# Astrophysical *S*-factor of the radiative proton capture on $^{14}$C at low energies

S.B. Dubovichenko[1,2,*], N. Burtebaev[2,†], A.V. Dzhazairov-Kakhramanov[1,2,‡], D.K. Alimov[2]

[1]*V. G. Fessenkov Astrophysical Institute "NCSRT" NSA RK, 050020, Observatory 23, Kamenskoe plato, Almaty, Kazakhstan*
[2]*Institute of Nuclear Physics CAE MINT RK, 050032, str. Ibragimova 1, Almaty, Kazakhstan*
[*]dubovichenko@mail.ru
[†]burteb@inp.kz
[‡]albert-j@yandex.ru

**Abstract:** The phase shift analysis for position location of the $^2S_{1/2}$ resonance at 1.5 MeV was carried out on the basis of the known experimental measurements of the excitation functions of the p$^{14}$C elastic scattering at four angles from 90° to 165° and more than 100 energy values in the range from 600–800 to 2200–2400 keV. Also, the possibility to describe the available experimental data on the astrophysical *S*-factor for the proton capture reaction on $^{14}$C to the ground state of $^{15}$N at astrophysical energies was considered in the frame of modified potential cluster model.

## 1. Introduction

Recently, for example in work [1], it was supposed that baryon number fluctuations in the early Universe lead to the formation of high-density proton-rich and low-density neutron-rich regions. This might be the result of the nucleosynthesis of elements with mass $A \geq 12$ in the neutron-rich regions of the early Universe [2,3]. The special interest is the nucleus $^{14}$C, which is produced by successive neutron capture

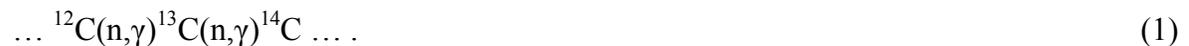

… $^{12}$C(n,γ)$^{13}$C(n,γ)$^{14}$C … .   (1)

The nucleus $^{14}$C has a half-life about 6000 years and is stable on the time scale of the Big-Bang nucleosynthesis. Therefore the synthesis of elements with mass $\geq 14$ depends on the rate of the neutron, alpha, and proton capture reactions on $^{14}$C. Because the cross section for neutron capture on $^{14}$C at thermal energies is very small (σ ≤ 1 μb [10]) and it is located at the level of 5–15 μb in the region of 100–1000 keV, it is assumed [2] that the alpha capture reaction is dominant. However, the proton capture on $^{14}$C might be of equal importance, because it depends on the proton abundance and density of their distribution in certain regions of the early Universe.

In addition, it can not be excluded that similar results can be dependent on the presence of dark energy and its concentration [11], on the growth of baryonic matter perturbations [12], or on the rotation of the early Universe [13]. However, perturbations in primordial plasma can not only stimulate the process of nucleosynthesis [14] but also kill it, for example, through the growth of the perturbations of nonbaryonic matter of the Universe [15] or because of the oscillations of cosmic strings [16].

Note furthermore that the results of new studies of the reaction $^{14}$C(p, γ)$^{15}$N in the nonresonance energy range [4] lead to the cross sections that higher on the order and more than the cross sections obtained earlier in work [17]. It allows one to obtain higher rate of the $^{14}$C(p, γ)$^{15}$N reaction at lower temperatures, notably, lower than 0.3T$_9$, that is essentially rise the role of this reaction for synthesis of heavier elements in the low energy range at the different stages of formation and development our Universe [2].

Now, take notice that the possibilities of simple two-body potential cluster model (PCM) were not completely studied up to now, particularly, if it use the conception of forbidden states (FSs) [18] and directly take into account the resonance behavior of elastic scattering phase shifts of interacting particles at low energies [19,20]. This model can be named as the modified PCM (MPCM with FS or simply MPCM). The MPCM with FS can be used, taking into account the classification of orbital states according to Young tableaux and also the resonance behavior of the elastic scattering phase shifts or resonances in spectra of the final nucleus for the considered initial channel. Such approach, as it was shown earlier [19-21], in many cases allows one to obtain quite adequate results in description of different experimental studies of the total cross sections for thermonuclear radiative capture processes.

Therefore, in continuing to study the processes of radiative capture [6,7], we will consider the p$^{14}$C → $^{15}$Nγ reaction within the framework of the MPCM at low energies. The potentials of intercluster interactions for scattering processes of the initial particles are constructed based on the reproduction of the elastic scattering phase shifts taking into account their resonance behavior or based on the structure of spectra of resonance states for the final nucleus in the initial channel. The intercluster potentials are constructed based on the description both of the binding energies of these particles in the final nucleus and of certain basic characteristics of these states [19-21] for the bound state (BS) or the ground state (GS) of nuclei, formed as a result of the capture reaction in the cluster channel, which coincide with the initial particles.

## 2. Structure of cluster states

Going to the analysis of the cluster states in the $N^{14}A$ system with the formation of the nuclei $^{15}$N or $^{15}$O in the GS, let us note that the classification of the orbital states of $^{14}$C in the n$^{13}$C channel or $^{14}$N in the p$^{13}$C channel according to Young tableaux was considered by us earlier in works [22-29]. We regard the results of the classification of the GSs of $^{15}$O and $^{15}$N by orbital symmetry in the considered channels as qualitative, because there are no complete tables of Young tableaux productions for systems with more than eight nucleons [30], which have been used in earlier similar calculations [19-21]. At the same time, simply based on such a classification, we succeeded in describing the available experimental data on the radiative capture of neutrons on $^{13}$C, $^{14}$C and $^{14}$N [22-25], and also capture of protons on $^{13}$C [26]. This is why the classification procedure by orbital symmetry given above was used here for the determination of the number of FSs and allowed states (ASs) in partial intercluster potentials and, consequently, to the specified number of nodes of the wave function of the relative motion of the clusters.

Furthermore, we will suppose that for $^{14}A$ it is possible to assume the orbital Young tableau in the form {4442} [22-26,31]; therefore, for the $N^{14}A$ system, we have {1} × {4442} → {5442} + {4443} in the frame of 1p-shell [30,32]. The first of the obtained tableaux is compatible with orbital moments $L = 0$, and 2, and is forbidden because it contains five nucleons in the s-shell. The second tableau is allowed and is compatible with orbital moments $L = 1$ [32]. As mentioned before, the absence of tables of Young tableaux productions for when the number of particles is 14 and 15 prevents the exact classification of the cluster states in the considered system of particles. However, qualitative estimations of the possible Young tableaux for orbital states allow us to detect the existence of the FS in the $^2S$ wave and the absence of FS for the $^2P$ states. The same structure of FSs and ASs in the different partial waves allows us to construct the potentials of intercluster interactions required for the calculations of the astrophysical $S$-factors, in this case for the proton radiative capture reaction on $^{14}$C.

Thus, by limiting our consideration to only the lowest partial waves with orbital moment $L = 0$, and 1, it could be said that for the n$^{14}$C system (for $^{14}$C it is known $J^\pi, T = 0^+, 1$), the forbidden and allowed states exist in the $^2S_{1/2}$ wave potential. The last of them corresponds to the GS of $^{15}$C with $J^\pi = 1/2^+$ and is at the binding energy of the n$^{14}$C system of -1.21809 MeV in the center-of-mass system (c.m.) [33]. At the same time the potentials of the $^2P$ waves of elastic scattering do not have FSs. Considering the p$^{14}$C system let us note that there is the FS in the potential of the $^2S_{1/2}$ wave, and in the $^2P_{1/2}$ wave there is only AS which corresponds to the GS of $^{15}$N with $J^\pi = 1/2^-$ and is at the binding energy of the p$^{14}$C system of -10.2074 MeV [33].

In the case of the n$^{14}$N system (for $^{14}$N we have $J^\pi, T = 1^+0$) the bound FS there is in the potentials of the $S$ scattering wave, and the $^2P_{1/2}$ wave has only the AS, which corresponds to the GS of $^{15}$N with $J^\pi = 1/2^-$ and is at the binding energy of the n$^{14}$N system of -10.8333 MeV [33]. For the p$^{14}$N system we obtain the similar results – there is the FS in the potentials of the $S$ scattering wave, and the $^2P_{1/2}$ wave has only the AS, which corresponds to the GS of $^{15}$O with $J^\pi = 1/2^-$ and is at the binding energy of the p$^{14}$N system of -7.2971 MeV [33].

Now let us consider the whole spectrum of resonance states in the p$^{14}$C system, i.e., states at positive energies. There are no resonance levels at the energies lower than 1 MeV in the spectra of $^{15}$N for the p$^{14}$C channel, which would have the width value more than 1 keV, however at 1.5 MeV in the laboratory system (l.s.) the very wide resonance at $J^\pi = 1/2^+$ with the width of 405 keV in c.m. is observed. Furthermore, the $^2S_{1/2}$ potential with the FS that describes this resonance will be constructed, and the potentials of the $^2P$ scattering waves can be equalized to zero, because they have no FSs. As it was mentioned before, the GS of $^{15}$N in the p$^{14}$C channel is the $^2P_{1/2}$ wave, therefore it is possible to consider the $E1$ transition from the resonance $^2S_{1/2}$ scattering wave at 1.5 MeV to the doublet $^2P_{1/2}$ GS of $^{15}$N:

1. $^2S_{1/2} \rightarrow\, ^2P_{1/2}$.

The $M1$ transitions to the GS from the doublet $^2P$ scattering waves, which potentials simply equal to zero, are possible in principle

2. $^2P_{1/2} \rightarrow\, ^2P_{1/2}$
   $^2P_{3/2} \rightarrow\, ^2P_{1/2}$

And also the $E1$ transition from the $^2D_{3/2}$ wave to the $^2P_{1/2}$ GS

3. $^2D_{3/2} \rightarrow\, ^2P_{1/2}$.

However, as it was obtained later as a result of our calculations, the transition 2 and 3, in comparison with the process 1, lead to very small cross sections and they are not taken into account in future calculations.

At first, carry out the standard phase shift analysis of the differential cross sections of the elastic p$^{14}$C scattering, because all intercluster potentials for describing of the proton radiative capture process on $^{14}$C will be based on the phase shifts of the elastic p$^{14}$C scattering.

## 3. Methods of calculation

Let us give firstly the basic expressions for the phase shift analysis of the p$^{14}$C elastic scattering and note that earlier we already have performed the phase shift analysis in systems p$^6$Li [34], n$^{12}$C [35], p$^{12}$C [36], $^4$He$^4$He [37], $^4$He$^{12}$C [38], p$^{13}$C [39], and n$^{16}$O [40], meanwhile, essentially at astrophysical energies [21].

The nuclear scattering phases obtained on the basis of the experimental differential cross sections allow one to extract certain information about the structure of resonance states of light atomic nuclei [41]. In this case, the processes of the elastic scattering of particles with total spin of 1/2 on the nucleus with zero spin take place in nuclear systems like $N^4$He, $^3$H$^4$He, $N^{12}$C, $N^{16}$O, $N^{14}$C etc. The cross section of the elastic scattering of such particles is presented in the simple form [42]

$$\frac{d\sigma(\theta)}{d\Omega} = |A(\theta)|^2 + |B(\theta)|^2, \qquad (2)$$

where

$$A(\theta) = f_c(\theta) + \frac{1}{2ik}\sum_{L=0}^{\infty}\{(L+1)S_L^+ + LS_L^- - (2L+1)\}\exp(2i\sigma_L)P_L(\cos\theta),$$

$$B(\theta) = \frac{1}{2ik}\sum_{L=0}^{\infty}(S_L^+ - S_L^-)\exp(2i\sigma_L)P_L^1(\cos\theta),$$

$$f_c(\theta) = -\left(\frac{\eta}{2k\sin^2(\theta/2)}\right)\exp\{i\eta\ln[\sin^{-2}(\theta/2)] + 2i\sigma_0\}. \qquad (3)$$

Here $S_L^\pm = \eta_L^\pm \exp(2i\delta_L^\pm)$ – scattering matrix, $\delta_L^\pm$ – required scattering phase shifts, $\eta_L^\pm$ – inelasticity parameters, and signs "±" correspond to the total moment of system $J = L \pm 1/2$, $k$ – wave number of the relative motion of particles $k^2 = 2\mu E/\hbar^2$, $\mu$ – reduced mass, $E$ – the energy of interacting particles in the center-of-mass system, $\eta$ – Coulomb parameter.

The multivariate variational problem of finding these parameters at the specified range of values appears when the experimental cross sections of scattering of nuclear particles and the mathematical expressions, which describe these cross sections with certain parameters $\delta_L^J$ – nuclear scattering phase shifts, are known. Using the experimental data of differential cross-sections of elastic scattering, it is possible to find a set of phase shifts $\delta_L^J$, which can reproduce the behavior of these cross-sections with certain accuracy. Quality of description of experimental data on the basis of a certain theoretical function or functional of several variables (2,3) can be estimated by the $\chi^2$ method, which is written as

$$\chi^2 = \frac{1}{N}\sum_{i=1}^{N}\left[\frac{\sigma_i^t(\theta) - \sigma_i^e(\theta)}{\Delta\sigma_i^e(\theta)}\right]^2 = \frac{1}{N}\sum_{i=1}^{N}\chi_i^2, \quad (4)$$

where $\sigma^e$ and $\sigma^t$ are experimental and theoretical, i.e., calculated for some defined values of the scattering phase shifts cross-sections of the elastic scattering of nuclear particles for *i*-angle of scattering, $\Delta\sigma^e$ – the error of experimental cross-sections at these angles, $N$ – the number of measurements. The details of the using by us searching method of scattering phase shifts were given in [42] and in our works [21,37].

Furthermore, we will determine astrophysical *S*-factors, which characterize behavior of the total cross section of the nuclear reaction at the vanishing energy, and they have the following form [17]

$$S(NJ, J_f) = \sigma(NJ, J_f) E_{cm} \exp\left(\frac{31.335 Z_1 Z_2 \sqrt{\mu}}{\sqrt{E_{cm}}}\right), \quad (5)$$

where $\sigma$ is the total radiative capture cross-section (barn), $E_{cm}$ is the center of mass energy of particles (keV), $\mu$ is the reduced mass of particles in the initial channel (atomic mass unit), $Z_{1,2}$ are the charges of particles in units of elementary charge, and $N$ is the $E$ or $M$ transitions of the $J$ multipole ordered from the initial $J_i$ to the final $J_f$ nucleus state. The numerical coefficient 31.335 was obtained on the basis of up-to-date values of fundamental constants [43].

The total radiative capture cross sections $\sigma(NJ, J_f)$ for the *EJ* and *MJ* transitions in the case of the PCM are given, for example, in work [44] or [19-25,45-47] and they have the following form:

$$\sigma_c(NJ, J_f) = \frac{8\pi K e^2}{\hbar^2 q^3} \frac{\mu}{(2S_1+1)(2S_2+1)} \frac{J+1}{J[(2J+1)!!]^2}$$
$$\times A_J^2(NJ, K) \sum_{L_i, J_i} P_J^2(NJ, J_f, J_i) I_J^2(J_f, J_i) \quad (6)$$

where $\sigma$ – total radiative capture cross section; $\mu$ – reduced mass of initial channel particles; $q$ – wave number in initial channel; $S_1$, $S_2$ – spins of particles in initial channel; $K$, $J$ – wave number and momentum of $\gamma$-quantum in final channel.

The value $P_J$ for electric orbital *EJ(L)* transitions ($S_i = S_f = S$) has the form [19-25,44]

$$P_J^2(EJ, J_f, J_i) = \delta_{S_i S_f}\left[(2J+1)(2L_i+1)(2J_i+1)(2J_f+1)\right](L_i 0 J 0 | L_f 0)^2 \begin{Bmatrix} L_i & S & J_i \\ J_f & J & L_f \end{Bmatrix}^2,$$

$$A_J(EJ, K) = K^J \mu^J \left(\frac{Z_1}{m_1^J} + (-1)^J \frac{Z_2}{m_2^J}\right), \quad I_J(J_f, J_i) = \langle \chi_f | R^J | \chi_i \rangle. \quad (7)$$

Here, $S_i$, $S_f$, $L_f$, $L_i$, $J_f$, and $J_i$ – total spins, angular and total moments in initial (*i*) and final (*f*) channels; $m_1$, $m_2$, $Z_1$, $Z_2$ – masses and charges of the particles in initial channel; $I_J$ –integral over wave functions of initial $\chi_i$ and final $\chi_f$ states, as functions of cluster relative motion with intercluster distance $R$.

Stop now in more detail treatment of the procedure and methods of construction using in the similar calculations of partial potentials at the given orbital moment $L$, having estimated criteria and sequence of parameter's finding and having noted to their errors and ambiguities [19-25]. In the first place, it is possible to find potential parameters of the BSs, which, at the given number of the allowed and forbidden states in this partial wave, are fixed quite unambiguously by the binding energy, the charge radius of nucleus and the asymptotic constant (AC) in the considered channel, for example, p$^{14}$C. In other words, the nucleus, in this case, $^{15}$N with 100% probability is considered like two-body p$^{14}$C system, and therefore, the spectroscopic factor $S_f$-factor of this channel simply equal to unit. The accuracy of the obtaining of parameters of the BS potential is connected, in the first place, with the AC accuracy, which usually equals 10–20%. There are no another ambiguities in this potential, because the state classification according to Young tableaux allows one to fix explicitly the number of the BSs, forbidden or allowed in the given partial wave, which completely determine its depth, and width of the potential depends wholly from the AC value. Let us note that for the considered here p$^{14}$C system we failed to find the AC values, obtained earlier in the independent calculations.

The intercluster potential of the nonresonance scattering process by phase shifts at the given number of BSs, ASs, and FSs in the considered partial wave also constructs quite unambiguously. The accuracy of determination of the parameters of such potential is connected, in the first place, with the accuracy of the scattering phase shift extraction from the experimental data and may reach to 20–30%. And here this potential does not contain ambiguities, because the state classification according to Young tableaux allows one to fix the number of BSs that completely determines its depth unambiguously, and the width of potential at the specified depth is determined by the shape of phase shift. At the construction of the nonresonance scattering potential according to the nuclear spectrum data in the certain channel, it is difficult to estimate the accuracy of finding the potential parameters even at the given BSs, although it is possible, apparently, to hope that it is a little more than the error in the previous case. This potential, as usual suggested for the energy range up to 1 MeV, should to lead to the scattering phase shift approximated to zero or gives smoothly dropped shape of the phase shift, if there are no resonance levels in nuclear spectra [19-25].

At the analysis of the resonance scattering the potential is constructed completely unambiguously when in the considered partial wave at the energies 1–2 MeV there is the resonance. At the given number of BSs its depth unambiguously fixes by the resonance energy of the level, and the width is formed by the width of such resonance. The error of its parameters usually does not exceed the error of determination of the width of such level and usually equals 3–5%. Meanwhile, this concerns to the construction of the partial potential according to the scattering phase shifts and to the determination of its parameters by the resonance in spectra of final nucleus.

Consequently, all potentials do not contain ambiguities inherent in optical model [42], and allow correctly describe the total cross sections of the radiative capture processes without additional involvement such notion as spectroscopic factor $S_f$ – as it was said before, its value simply takes equal to unit. Here, it is possible to think that there is no necessity to use the additional factor $S_f$ [44] at the consideration capture reaction in the framework of the MPCM for the potentials matched in the continuous spectrum with the characteristics of scattering processes, which take into account the resonance shape of phase shifts and discrete spectrum, describing the basic characteristics of the BSs. Evidently, all effects that take part into reaction, including probability of cluster configuration, are taken into account at similar construction of the interaction potentials. It is possible because the potentials are constructed taking into account the structure of the FSs and on the basis of description of the observed, i.e., the experimental characteristics of interacting clusters in the initial channel and the certain nucleus formed in the final state at the description of it by the cluster structure, consisting with initial particles. Thus, the existence of the $S_f$, evidently already takes into account in the wave functions of the BSs of clusters, determined on the basis of such potentials at solving the Schrödinger equation [19-25].

The next values of particle masses are used in the given calculations: $m_p$ = 1.00727646577 amu [48], and $m(^{14}C)$ = 14.003242 amu [49], and constant $\hbar^2/m_0$ is equal to 41.4686 MeV fm$^2$. The Coulomb parameter $\eta = \mu Z_1 Z_2 e^2/(q\hbar^2)$ was represented as $\eta = 3.44476\, 10^{-2} Z_1 Z_2 \mu/q$, where $q$ is the wave number determined by the energy of interacting particles in the initial channel (in fm$^{-1}$). The Coulomb potential with $R_{Coul.}=0$ was represented as $V_{Coul.}$(MeV) = 1.439975 $Z_1Z_2/R$, where $R$ is the relative distance between the initial channel particles (fm).

## 3. Radiative proton capture on $^{14}$C in cluster model
### 3.1. Phase shifts and potentials of the elastic scattering

The excitation functions from the work [50], measured at 90°, 125°, 141° and 165° in the energy range from 0.6 to 2.3 MeV (l.s.), are shown in Figs. 1a,b,c,d by dots. These data are used by us further for carrying out of the phase shift analysis and extracting of the resonance form of the $^2S_{1/2}$ scattering phase at 1.5 MeV. The results of the present analysis are shown in Figs. 2a,b,c,d by dots, and the solid lines in Fig. 1 show cross sections calculated with the obtained scattering phase shifts. About 120 first points, cited in work [50], in the stated above energy range were used in this analysis. In addition, it was obtained that for description of the cross sections in the excitation functions, at least at the energies up to 2.2–2.3 MeV, there is no need to take into account $^2P$ or $^2D$ scattering waves, i.e., their values simply can be equalized to zero. Since only one point in the cross sections of excitation functions is considered for each energy and angle, therefore the value of $\chi^2$ at all energies and angles usually lays at the level $10^{-2}$–$10^{-10}$ and taking into account another partial scattering phase shifts already does not lead to its decrease.

The resonance energy, as it is seen in Fig. 2, obtained from the excitation function at 90° is at the interval of 1535–1562 keV for which the phase shift value lays within limits of 87°–93° with the value of 90° at 1554 keV. The resonance energy obtained from the excitation function at 125° is at the interval of 1551–1575 keV for which the phase shift value lays within limits of 84°–93°. The resonance energy obtained from the excitation function at 141° is at the interval of 1534–1611 keV for which the phase shift value lays within limits of 84°–90° with the value of 90° at 1534, 1564 and 1611 keV. The resonance energy obtained from the excitation function at 165° is at the interval of 1544–1563 keV for which the phase shift value lays within limits of 87°–91°. For so noticeable scatter of values, it can be said only that the resonance value lays within limits of 1534–1611 that is, in general, agree with data [33], where the resonance energy value of 1509 keV is given.

Let us note that work [50] mentions that the detailed analysis of the resonances, including 1.5 MeV, was not carried out, because it was done earlier in works [51,52] on the basis of the proton capture reaction on $^{14}$C. Here, as one can see from the results of the phase shift analysis, the resonance energy at 1.5 MeV slightly overestimated. However, as it was seen in Fig. 1, the data spread on cross sections in excitation functions is too large for clear conclusion about the energy of resonance. Apparently, the additional and more modern measurements of the elastic scattering cross sections are required, in order to on the basis of these data to perform more unambiguous conclusion about the resonance energy at 1.5 MeV.

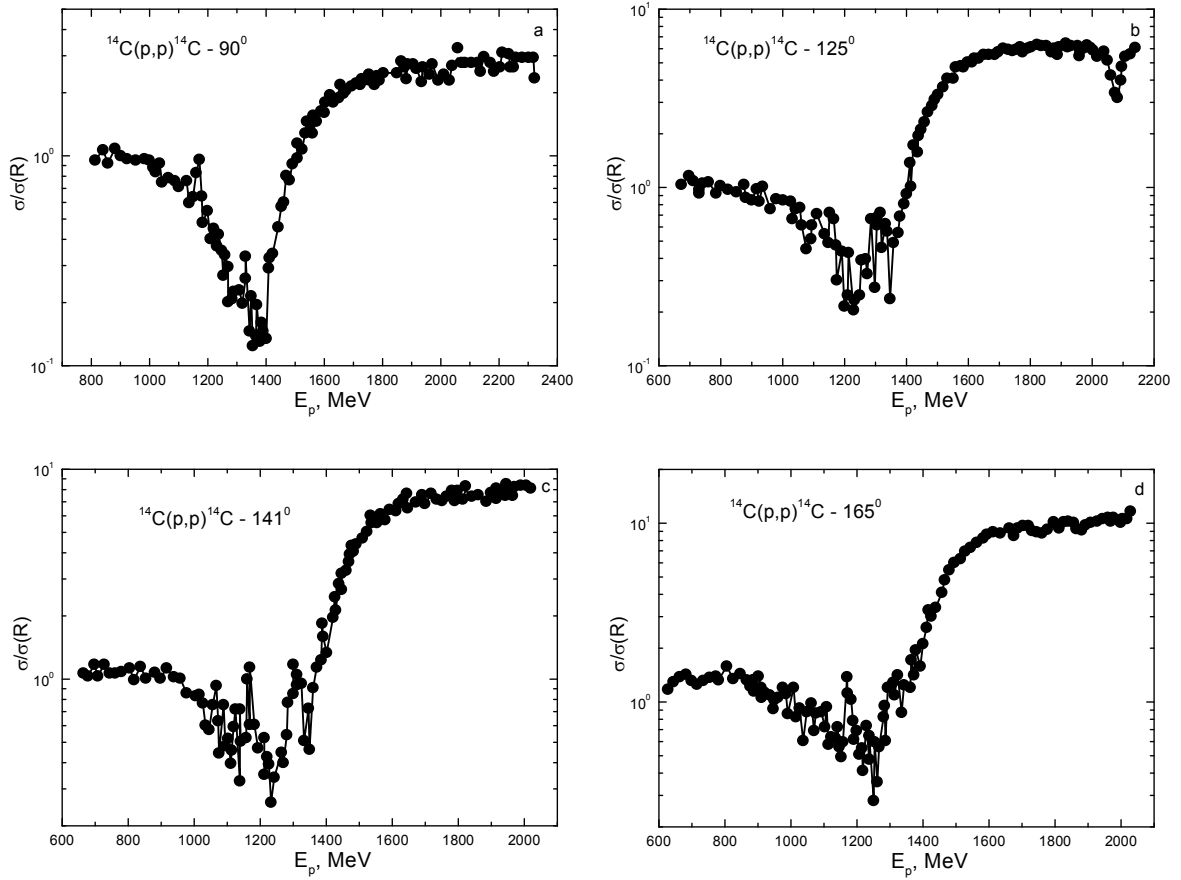

**Figs. 1a,b,c,d.** The excitation functions in the elastic p$^{14}$C scattering in the range of the $^2S_{1/2}$ resonance [50]. The solid line – their approximation on the basis of the obtained scattering phase shifts.

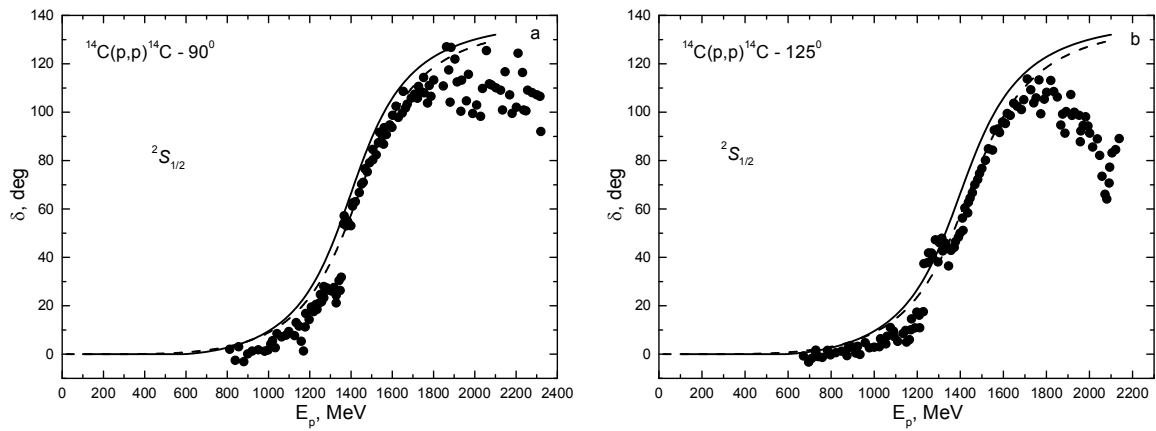

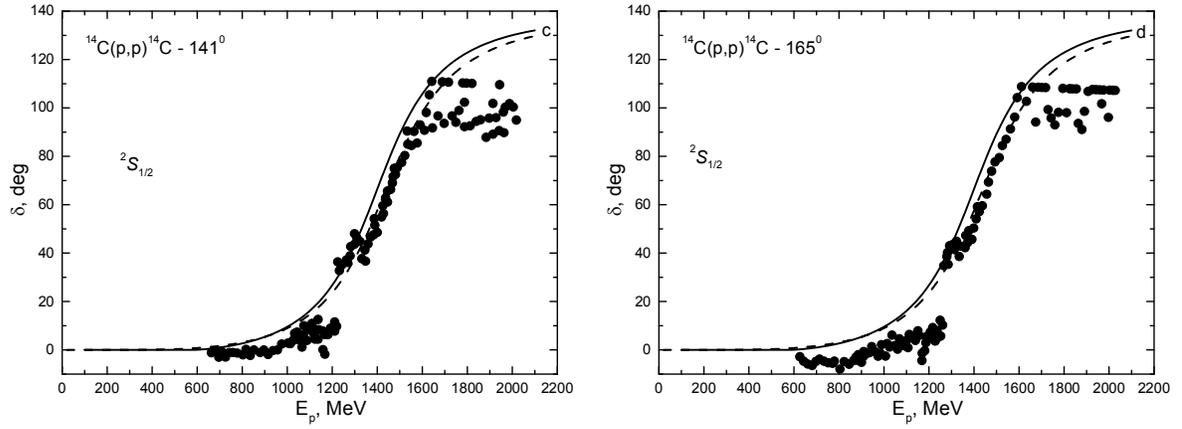

**Figs. 2a,b,c,d.** The $^2S_{1/2}$ elastic phase shift of the p$^{14}$C scattering at low energies, obtained on the basis of the excitation functions, shown in Figs. 1a,b,c,d. Points – results of our phase shift analysis, carried out on the basis of data from [50], lines – calculation of the phase shift with the potentials given in the text.

For description of the obtained $^2S_{1/2}$ scattering phase in the phase shift analysis it is possible to use simple Gaussian potential of the form:

$$V(r) = -V_0 \exp(-\alpha r^2) \qquad (8)$$

with FSs and parameters

$$V_0 = 5037.0 \text{ MeV}, \quad \alpha = 12.0 \text{ fm}^{-2}, \qquad (9)$$

which lead to the scattering phase shifts with the resonance at 1500 keV (l.s.) and with the width of 530 keV (c.m.) that is in a good agreement with the available experimental data [33]. The parameters of this potential were matched to reproduce in general the resonance data exactly from [33], which were obtained in works [51,52]. The phase shift of this potential is shown in Figs. 2a,b,c,d by the solid lines and at the resonance energy reaches the value of 90(1)°. The energy behavior of the scattering phase shift of this potential correctly describes obtained in the phase shift analysis scattering phases in whole, taking into account the shift of the resonance energy approximately at 30–50 keV relative to results [33]. The calculated phase shift line for this potential is parallel to points, obtained in our phase shift analysis, for all scattering angles.

For more accurate description of the obtained in the phase shift analysis data the next potential is needed

$$V_0 = 5035.5 \text{ MeV}, \quad \alpha = 12.0 \text{ fm}^{-2}. \qquad (10)$$

It leads to the resonance energy of 1550 keV, its width of 575 keV, and the calculation results of the $^2S_{1/2}$ phase shift are shown in all Figs. 2 by the dashed line. As one can see this line appreciably better reproduce results of the carried out here phase shift analysis.

It should be noted over again that the potential is constructed completely unambiguously, if the number of FSs is given (in this case, it is equal to unit), according to the known energy of the resonance level in spectra of any nucleus [33] and its width. In other words, it is not possible to find another combination of the parameters $V$ and $\alpha$, which could be possible to describe the resonance energy of level and its width correctly. The depth of such potential unambiguously determines the resonance location, i.e., resonance energy of the level, and its width α specifies the certain width of this resonance state, which have to correspond to experimental observable values [33].

For the $^2P_{1/2}$ GS potential of $^{15}$N without FSs in the cluster p$^{14}$C-channel the following parameters were found

$$V_0 = 221.529718 \text{ MeV}, \quad \alpha = 0.6 \text{ fm}^{-2}. \tag{11}$$

It allows to find the value of $R_m = 2.52$ fm for the mass radius, the value of $R_{ch} = 2.47$ fm for the charge radius, the binding energy of -10.207400 MeV at the accuracy of the finite-difference method [53] of $10^{-6}$ MeV. The phase shift of such potential decreases smoothly and at 2 MeV approximately equals 179°, and for the asymptotic constant in the dimensionless form [54]

$$\chi_L(r) = \sqrt{2k_0} C_w W_{-\eta L+1/2}(2k_0 r), \tag{12}$$

the value of 1.80(1) was obtained at the distance interval 3–10 fm. The mentioned AC error is obtained by its averaging over the given distance interval. The value of 2.56(5) fm [33] was used for radius of $^{14}$C, for radius of $^{15}$N it is known the value of 2.612(9) fm [33], the proton radius is equal to 0.8775(51) [48]. Note once more that we failed in finding AC data in this channel, obtained in another works by independent methods.

### 3.2. The total capture cross sections

Going to the description of the obtained results, let us note that the experimental data for total cross section of the proton radiative capture on $^{14}$C to the GS of $^{15}$N or for the astrophysical *S*-factor were measured in work [4] for the energy range 260-740 keV – these results are shown by squares in Figs. 3 and 4 and in numerical form were taken by us from the data base [55]. The capture cross section to the GS with potential (11) for the *E*1 transition from the resonance $^2S$ scattering wave with potential (9) was considered for their description. The calculation results of the total cross section and the astrophysical *S*-factor are shown in Figs. 3 and 4 by the solid lines, respectively. They practically completely describe the experimental data at all measured in work [4] energies. Meanwhile, it is good to see that in Fig. 4 the calculated line is in the band of experimental errors and ambiguities of available measurements [4].

It should be noted here that if scattering potential (9) was constructed exclusively based on the characteristics of the resonance at 1.5 MeV, then because of the absence of the AC value in the p$^{14}$C channel the GS potential (11) is determined so that to describe available experimental data of the total capture cross sections [4] to the best advantage. However, this potential leads to the quite reasonable radius value of $^{15}$N, and the AC value obtained with it could be used further for comparison with the results of other works. Moreover, the studies of such capture reactions in the used model allow one to extract the AC value, though its accuracy hardly exceeds 10-20%.

The calculated values of the *S*-factor at the resonance energy also can be used in future for comparison with modern measurements of this reaction. But the measurement spread on the *S*-factor [4] and the experimental errors of these measurements, for example, in the range 258-260 keV reach 30%. Therefore, the calculation accuracy of the maximum *S*-factor value, which in these calculations equals 81-82 keV b at $E_p$=1390keV, will be approximately the same, although the phase shift resonance of potential (9) is at 1500 keV. The value of the calculated *S*-factor remains almost constant and equals 4.5(1) keV b in the region of low energies, notably 100-140 keV. Apparently, just that very value can be considered as the *S*-factor at zero energy for the considered here proton radiative capture reaction on $^{14}$C.

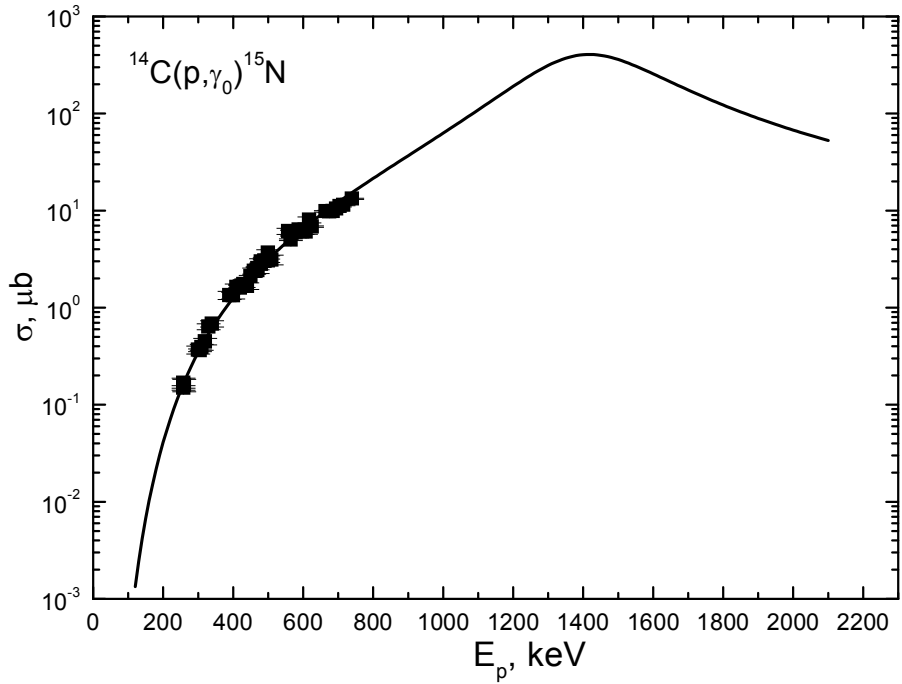

Fig. 3. The total cross section of the proton capture on $^{14}$C to the GS of $^{15}$N. The experimental data are from work [4].

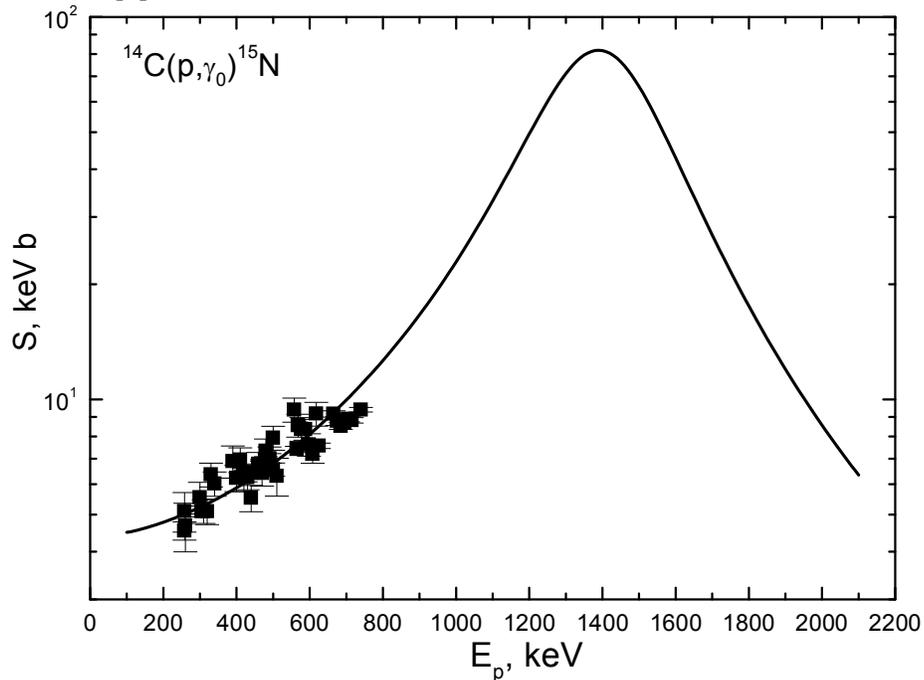

Fig. 4. The astrophysical *S*-factor of the proton capture on $^{14}$C to the GS of $^{15}$N. The experimental data are from work [4].

## 4. Conclusion

Thereby, the resonance $^2S_{1/2}$ phase shift of the p$^{14}$C elastic scattering at energies from 0.6 MeV to 2.3 MeV was found as a result of the carried out phase shift analysis of the experimental differential cross sections in excitation functions [50]. The resonance energy of the phase shift is in a quite agreement with the level spectrum of $^{15}$N [33] in the p$^{14}$C channel. The results of the carried out phase shift analysis, i.e., phase shift of the elastic p$^{14}$C scattering and the data of resonances of $^{15}$N [33], allow one to parametrize intercluster interaction potentials for scattering processes in the resonance $^2S_{1/2}$ wave. These potentials, by-turn, can be used further for carrying out certain calculations for different astrophysical problems, partially considered, for example, in [19-21].

It should be noted that it is not enough data of work [50] for carrying out of the accurate phase shift analysis and the another additional data on differential cross sections, for example, angular distributions in the resonance region are needed. Such data can allow one to determine the location of the $^2S_{1/2}$ resonance in the region of 1.5 MeV and more accurately determine its width exactly on the basis of the elastic scattering phase shifts.

Furthermore, the measured cross section or the astrophysical S-factor of the proton capture reaction on $^{14}$C are succeeded to correctly describe on the basis only of the E1 transition from the resonance $^2S_{1/2}$ scattering wave with FS to the $^2P_{1/2}$ GS without FS of $^{15}$N, considered in the two-body p$^{14}$C model. The carrying out in future more detailed measurements of total cross sections of this reaction, especially, in the resonance region at 1.5 MeV allows one to draw, apparently, more concrete conclusions about the quality of description of the considered cross sections of the reaction of the proton capture on $^{14}$C in the framework of the MPCM.

Let us note that this is already twenty seventh cluster system from the considered by us earlier based on the MPCM with the classification of orbital states according to Young tableaux, where it is possible to obtain acceptable results for description of the characteristics of the radiative capture processes of nucleons and light nuclei on nuclei of 1p-shell. The properties of these cluster nuclei, their characteristics and considered cluster channels are given in Table 1. The last results obtained in the framework of the MPCM are given in works [26-29,31,56-64].

**Table 1.** The characteristics of nuclei and cluster systems, and references to works in which they were considered.[+)]

| No. | Nucleus ($J^\pi$,T) | Cluster channel | $T_z$ | T | Refs. |
|---|---|---|---|---|---|
| 1. | $^3$He (1/2$^+$,1/2) | p$^2$H | +1/2 + 0 = +1/2 | 1/2 | [19,27] |
| 2. | $^3$H (1/2$^+$,1/2) | n$^2$H | -1/2 + 0 = -1/2 | 1/2 | [21-25] |
| 3. | $^4$He (0$^+$,0) | p$^3$H | +1/2 - 1/2 = 0 | 0 + 1 | [19,56] |
| 4. | $^6$Li (1$^+$,0) | $^2$H$^4$He | 0 + 0 = 0 | 0 | [19,57] |
| 5. | $^7$Li (3/2$^-$,1/2) | $^3$H$^4$He | -1/2 + 0 = -1/2 | 1/2 | [19,57] |
| 6. | $^7$Be (3/2$^-$,1/2) | $^3$He$^4$He | +1/2 + 0 = +1/2 | 1/2 | [19,57] |
| 7. | $^7$Be (3/2$^-$,1/2) | p$^6$Li | +1/2 + 0 = +1/2 | 1/2 | [19,46,47] |

| 8. | $^7$Li (3/2$^-$,1/2) | n$^6$Li | -1/2 + 0 = -1/2 | 1/2 | [21-25] |
|---|---|---|---|---|---|
| 9. | $^8$Be (0$^+$,0) | p$^7$Li | +1/2 - 1/2 = 0 | 0 + 1 | [19,56] |
| 10. | $^8$Li (2$^+$,1) | n$^7$Li | -1/2 - 1/2 = -1 | 1 | [21-25,28] |
| 11. | $^{10}$B (3$^+$,0) | p$^9$Be | +1/2 - 1/2 = 0 | 0 + 1 | [19,46,47] |
| 12. | $^{10}$Be (0$^+$,1) | n$^9$Be | -1/2 - 1/2 = -1 | 1 | [21-25,58] |
| 13. | $^{11}$C (3/2$^-$,1/2) | p$^{10}$B | 1/2 + 0 = 1/2 | 1/2 | [64] |
| 14. | $^{11}$B (3/2$^-$,1/2) | n$^{10}$B | -1/2 + 0 = -1/2 | 1/2 | [58] |
| 15. | $^{12}$C (0$^+$,0) | p$^{11}$B | +1/2 - 1/2 = 0 | 0 | [60] |
| 16. | $^{12}$B (1+,1) | n$^{11}$B | -1/2 - 1/2 = -1 | 1 | [61] |
| 17. | $^{13}$N (1/2$^-$,1/2) | p$^{12}$C | +1/2 + 0 = +1/2 | 1/2 | [19,29,31] |
| 18. | $^{13}$C (1/2$^-$,1/2) | n$^{12}$C | -1/2 + 0 = -1/2 | 1/2 | [21-25] |
| 19. | $^{14}$N (1$^+$,0) | p$^{13}$C | +1/2 - 1/2 = 0 | 0 + 1 | [26,29,31] |
| 20. | $^{14}$C (0$^+$,1) | n$^{13}$C | -1/2 – 1/2 = -1 | 1 | [21-25] |
| 21. | $^{15}$N (1/2$^-$,1/2) | p$^{14}$C | +1/2 – 1 = -1/2 | 1/2 | Present work |
| 22. | $^{15}$C (1/2$^+$,3/2) | n$^{14}$C | -1/2 – 1 = -3/2 | 3/2 | [21-25] |
| 23. | $^{15}$N (1/2$^-$,1/2) | n$^{14}$N | -1/2 + 0 = -1/2 | 1/2 | [21-25] |
| 24. | $^{16}$O (0$^+$,0) | p$^{15}$N | +1/2 - 1/2 = 0 | 0 | [62] |
| 25. | $^{16}$N (2$^-$,1) | n$^{15}$N | -1/2 - 1/2 = -1 | 1 | [21-25] |
| 26. | $^{16}$O (0$^+$,0) | $^4$He$^{12}$C | 0 + 0 = 0 | 0 | [19,63] |
| 27. | $^{17}$O (5/2$^+$, 1/2) | n$^{16}$O | -1/2 + 0 = -1/2 | 1/2 | [21-25] |

+) $T$ – isospin and $T_z$ – its projection, $J^\pi$ – total moment and parity.

Thus, the MPCM again confirms its ability correctly describe the cross sections of the processes like radiative capture of neutral and charged particles on light nuclei at thermal and astrophysical energies, using for this simple principles of classifications of FSs and ASs in two-cluster systems [19-25,29,31,46,47].

## Acknowledgments

The work was performed under the grant No. 0601/GF "The experimental and theoretical study of properties of the excited halo-states of neutron-rich nuclei $^9$Be, $^{11}$B, $^{13,15}$C and $^{15}$N" of the Ministry of Education and Science of the Republic of Kazakhstan.